\documentclass[
    ,final            
  ]
  {aipproc}

\layoutstyle{6x9}

\usepackage{natbib}
\usepackage{latexsym}
\usepackage{amssymb,amsmath}

\newcommand{\includefigurevh}[4]{\newpage {\centering
   \vspace*{#3}    \hspace*{#4}  \includegraphics*[scale=#1]{#2}}}

\newcommand{\refl}[1]{(\ref{#1})}
\newcommand{\eeql}[1]{\label{#1}\eeq}
\newcommand{\vp}{\ensuremath{\phi}}
\newcommand{\lag}{\ensuremath{\mathcal{L}}}
\newcommand{\ord}[1]{\ensuremath{\mathcal{O}(#1)}}

\newcommand{\ra}{\ensuremath{\rightarrow}}

\newcommand{\zpr}{\ensuremath{Z'}}
\newcommand{\mzp}{\ensuremath{M_{Z'}}}

\newcommand{\uprm}{\ensuremath{U(1)'}}

\newcommand{\skipblk}[1]{}                                                      
\def\bqa{\begin{eqnarray}}                                                      
\def\eqa{\end{eqnarray}}

\newcommand{\douba}[2]{\ensuremath{                                                  
\left( \begin{array}{c} #1    \\ #2 
        \end{array}\right)}}

\newcommand{\sto}{\ensuremath{SU(2) \x U(1)}}                                       
\newcommand{\st}{\ensuremath{SU(2)}}

\newcommand{\x}{\ensuremath{\times}}

\newcommand{\sinn}{\ensuremath{\sin^2\theta_W\,}}

\newcommand{\beq}{\begin{equation}}                                             
\newcommand{\eeq}{\end{equation}}     
                                                                                   
\newcommand{\oh}{\ensuremath{\frac{1}{2}}}

\usepackage{ifthen}
\newboolean{hyperlinks}
\setboolean{hyperlinks}{false}
\setboolean{hyperlinks}{true}
\ifthenelse{\boolean{hyperlinks}}{\usepackage{hyperref}\bibliographystyle{utphys_mod} 
}{
\bibliographystyle{aipproc} }

\begin{document}

\title{The Physics of New \uprm\ Gauge Bosons}


\author{Paul Langacker}{
  address={Institute for Advanced Study
,Princeton, NJ 08540}
}

\begin{abstract}
Additional \zpr\ gauge bosons are predicted by a wide variety of extensions of the standard model (SM). Possibilities include TeV-scale bosons with electroweak coupling, very light bosons which nearly decouple from the standard model particles, and bosons which communicate with a quasi-hidden sector. A broad survey is given of the theoretical possibilities and of the physics implications for particle physics and cosmology. Several novel examples, including light  \zpr s suggested by PAMELA, Stueckelberg \zpr s, and  \zpr s associated with the mediation of supersymmetry breaking, are described.
\end{abstract}

\maketitle


\section{Motivations}
An additional \zpr\ gauge boson is one of the best motivated extensions of the standard model or
MSSM. String and grand unified theories often involve large underlying groups, or (in the case of
Type IIa constructions) promote $SU(n)$ groups to $U(n)$. In many cases, it is more difficult to break the
 \uprm\ generators than the non-abelian ones, implying that extra \zpr s may survive to low energies as accidental
 remnants of the symmetry breaking. 
 Also, many supersymmetric \uprm\ models provide a natural solution to the  $\mu$ problem. In that case both the
  \sto\ and \uprm\ breaking scales are set by 
SUSY breaking scale (unless the breaking is associated with a flat direction), so one naturally expects the
\zpr\ mass to be comparable to the electroweak scale, up to an order of magnitude or so.
Similarly, alternative electroweak models (such as left-right symmetry) or alternatives to the
 elementary Higgs mechanisms, e.g., dynamical symmetry breaking or Little Higgs models,
involve new TeV scale physics and extended gauge groups that often lead to new TeV-scale \zpr s.
Models in which the standard model gauge bosons can propagate in large and/or warped  extra dimensions
 involve {Kaluza-Klein excitations}, with $M\sim R^{-1}\sim 2 {\ \rm TeV }\x
(10^{-17} {\rm  cm }/R)$ in the large dimension case. Another aspect is that \zpr s may provide a weak coupling between the ordinary sector
of matter and other sectors associated with dark matter or with supersymmetry breaking. Finally, the existence of an extra \zpr, especially at the TeV scale, would have extensive implications for
collider physics and cosmology. In this talk I will describe selected recent developments. More general
treatments may be found in several recent reviews~\citep{Langacker:2008yv,Rizzo:2006nw,Leike:1998wr}.

\section{The Standard Model with Additional \uprm s}
The interactions of the photon ($A$), $Z$ (i.e., $Z^0_1$) and other flavor-diagonal neutral gauge bosons
with fermions  is
\beq -\lag_{NC}= \underbrace{eJ^\mu_{em} A_\mu+g_1 J^\mu_1 Z^0_{1 \mu}}_{SM}+ \sum_{\alpha=2}^{n+1} g_\alpha J^\mu_\alpha Z^0_{\alpha \mu}, \eeql{f1}
where $g_\alpha$ are the gauge couplings (with $g_1=g/\cos\theta_W$), and the currents are
\beq J^\mu_\alpha= \sum_i \bar f_i \gamma^\mu[\epsilon_L^{{\alpha}}(i)P_L+\epsilon_R^{{\alpha}}(i)P_R] f_i. \eeql{f2}
 $\epsilon_{L,R}^{{\alpha}}(i)$ are the $U(1)_\alpha$  charges of the left- and right-handed components of  fermion $f_i$, and the theory is chiral for $\epsilon_{L}^{{\alpha}}(i)\ne \epsilon_{R}^{{\alpha}}(i)$. 
 It is often convenient to instead
 specify the charges $Q_\alpha$ for  the left-chiral 
fermion $f_L$ {and}  and left-chiral antifermion  $f^c_L$,
\beq Q_{\alpha f}=\epsilon_L^{{\alpha}}(f) \quad  \qquad Q_{\alpha f^c}=-\epsilon_R^{{\alpha}}(f). \eeql{f3}
For example, the SM charges for the $u_L$ and $u^c_L$ are
$Q_{1u}=\oh - \frac{2}{3} \sinn$ and
$Q_{1u^c}=+ \frac{2}{3} \sinn$.
One can similarly define the $U(1)_\alpha$  charge of the scalar field $\vp$  as $Q_{\alpha\vp}$.

For a single extra \zpr, the $Z-\zpr$ mass matrix after symmetry breaking is
\beq M^2_{Z-Z'} =   \left(
\begin{array}{cc}
 M_{Z^0}^2 & \Delta^2\vspace*{2pt} \\  \Delta^2 & M_{Z'}^2
\end{array} \right). \eeql{f4}
If, for example, the symmetry breaking is due to an \st-singlet $S$ and two doublets
$\phi_u= \douba{\phi^0_u}{\phi^-_u}, \  \phi_d= \douba{\phi^+_d}{\phi^0_d}$,
then
\beq
\begin{split}
M_{Z^0}^2 =&  \frac{1}{4} g_1^2 (|\nu_u|^2+|\nu_d|^2)  , \qquad
 \Delta^2 =  \oh g_1 g_2 (Q_u |\nu_u|^2-Q_d |\nu_d|^2) \\
M_{Z'}^2 =& g_2^2( Q_u^2 |\nu_u|^2 + Q_d^2|\nu_d|^2+Q_S^2 |s|^2),
\end{split}
\eeql{f5}
where
\beq  \nu_{u,d} \equiv \sqrt{2} \langle  \phi^0_{u,d} \rangle, \qquad s=\sqrt{2}  \langle S \rangle, \qquad
\nu^2=(|\nu_u|^2+|\nu_d|^2)\sim (246 {\rm\ GeV})^2 .
\eeql{f6}
The physical mass eigenvalues are $M_{1,2}^2$, and the  mixing angle
 $\theta$ is given by
 $ \tan^2\theta ={(M_{Z^0}^2-M_1^2)}/{(M_2^2-M_{Z^0}^2)}$.
In the important special case  $M_{Z'} \gg (M_{Z^0}, |\Delta|)$ one finds
\beq \begin{split}
M_1^2 & \sim M_{Z^0}^2 - \frac{\Delta^4}{M_{Z'}^2}\ll M_2^2 \quad \qquad M_2^2 \sim M_{Z'}^2 \\
\theta & \sim -\frac{\Delta^2}{M_{Z'}^2}\sim C \frac{g_2}{g_1} \frac{M_1^2}{M_2^2}
\text{\ \ with \  \ } C=
2\left[  \frac{Q_u |\nu_u|^2-Q_d |\nu_d|^2}{|\nu_u|^2+|\nu_d|^2} \right].
\end{split} \eeql{f8}

We have so far implicitly assumed canonical kinetic energy terms for the $U(1)$ gauge bosons. However,  $U(1)$ gauge invariance allow a more general kinetic mixing~\citep{Holdom:1985ag},
\beq \lag_{kin}\ra-\frac{1}{4} F^{0\mu\nu}_1 F^0_{1\mu\nu}-\frac{1}{4} F^{0\mu\nu}_2 F^0_{2\mu\nu}
-\frac {\sin \chi}{2} F^{0\mu\nu}_1 F^0_{2\mu\nu}
\eeql{f9}
for $U(1)_1\x U(1)_2$.
Such terms are usually absent initially, but a (usually small) $\chi$ may be induced by loops, e.g., from nondegenerate heavy particles, in running couplings
if heavy particles decouple, or at the string level. The kinetic terms may be put in canonical form by the non-unitary transformation
\beq  \douba{Z^0_{1\mu}}{Z^0_{2 \mu}}= 
\left(
\begin{array}{cc} 1 &   -\tan \chi   \\ 0 &   1/\cos \chi\end{array}\right)
\douba{\hat Z^0_{1\mu}}{\hat Z^0_{2\mu}},
\eeql{f10}
where  the $\hat Z^0_{\alpha}$  may still undergo ordinary mass mixing, as in \refl{f4}.
The kinetic mixing has a negligible effect on masses for  $|M_{Z_1}^2| \ll |M_{Z_2}^2|$
and $|\chi|\ll 1$, but the current coupling to the heavier boson is shifted,
\beq  -\lag \ra  g_1 J^\mu_1\hat Z_{1 \mu}+  (g_2 J^\mu_2 -g_1 \chi J^\mu_1)\hat Z_{2 \mu} .
\eeql{f11}

\section{Models}
There are an enormous number of models (e.g.,~\citep{Langacker:2008yv}), distinguished by their gauge coupling $g_2$, their mass scale,
their charges $Q_2$, the exotic fermions needed for anomaly cancellation, kinetic mixing, possible couplings to hidden sectors, whether one assumes supersymmetry, etc, and there is no simple general parametrization. Some of the major classes include:
\begin{itemize}
\item ``Canonical''  TeV-scale models with electroweak strength couplings, including
the sequential $Z_{SM}$ model with the same couplings to fermions as the $Z$,
models based on $T_{3R}$ and $B - L$, $E_6$ models, and
minimal gauge unification models.
\item Those based on new TeV scale dynamics, such as Little Higgs models, un-unified models, dynamical symmetry breaking models, e.g., with strong $t\bar t$ coupling, etc.
\item Kaluza-Klein excitations of the SM gauge bosons in models with large and/or warped extra dimensions.
\item Models in which the \zpr\ is decoupled from some or all of the SM particles,
such as leptophobic, fermiophobic or weak coupling models. These may have a low scale or even massless \zpr.
\item Models in which the \zpr\ couples to a hidden sector, e.g., associated with dark matter or supersymmetry breaking. Such a \zpr\ may (almost) decouple from the SM particles and serve as
a weakly coupled ``portal'' to the hidden sector,
or may couple to both sectors, e.g., to mediate supersymmetry breaking.
\item Supersymmetric models  with a secluded or intermediate scale \zpr, e.g., associated with (approximately) flat directions, small Dirac $m_\nu$ from higher-dimensional operators, etc.
\item Models with family-nonuniversal couplings, leading to \zpr-mediated FCNC.
\item String derived models, which may be lead to \zpr\ coupled to $T_{3R},$ $B-L,$ or $E_6$
couplings, or may appear ``random''.
\item St\"uckelberg models~\citep{Stueckelberg:1900zz}, which allow a \zpr\ mass without 
spontaneous symmetry breaking.
\item Anomalous \uprm\ models, which may be relevant to some string theories involving large dimensions.
\end{itemize}

\section{Recent developments}
There has been much recent work on models in which a \zpr\ connects the SM particles to an otherwise  hidden or dark sector. In some cases, the \uprm\ mainly couples to the
hidden sector, with a weak connection to the SM via small kinetic mixing (for a review, see~\citep{Goodsell:2009xc}), higher-dimensional operators, $D$-terms, Higgs fields, Chern-Simons $\zpr Z\gamma$ couplings, etc., leading to the possibility of invisible \zpr\ decays to hidden sector particles
and/or suppressed decays to the SM particles. Alternatively, the \uprm\ may couple directly to both sectors (as is common in string constructions), such as in \zpr\ mediation models.
Here, we mention several examples.

\subsection{A massless \zpr}
An interesting possibility is that there are two massless gauge bosons, the photon ($A$) and a new \zpr, presumably with no direct couplings to the SM sector.
A small kinetic mixing of the form \refl{f9} could induce the interaction
\beq  -\lag \ra e( J^{em} -\frac{g_2 \chi}{e} J^\mu_2)A_{ \mu} + g_2 J^\mu_2 Z'_{ \mu} ,
\eeql{f12}
which differs from \refl{f11} since both gauge bosons are massless.
In \refl{f12}, $g_2 J_2$ describes possible interactions of the \zpr\ with a hidden sector. 
For small $\chi$, e.g., $\chi=\ord{10^{-3}}$ the kinetic mixing induces effective
milli-electric charges for the hidden sector particles~\citep{Holdom:1985ag}.
Alternatively, the sectors could be connected by higher-dimensional operators~\citep{Dobrescu:2004wz},
leading, e.g., to $\mu\ra e Z'$. Implications of such models for dark matter have been studied in~\citep{Huh:2007zw,Feng:2009mn}, possible string origins in~\citep{Goodsell:2009xc,Abel:2008ai}, and  recent experimental and cosmological constraints
in~\citep{Afanasev:2008fv,Ehret:2009sq,Burrage:2009yz}.

\subsection{An MeV/GeV \zpr}
A related possibility is that a \zpr\ in the MeV-GeV range (sometimes referred to as a $U$-boson~\citep{Fayet:2006sp,Boehm:2003hm}) acquires a weak coupling to $J_{em}$ by kinetic mixing.
Such a particle could have implications for or is constrained by, e.g., 
$g_\mu-2$, $e^+e^-\ra U\gamma\ra e^+e^-\gamma$, and even the HyperCP events~\citep{Borodatchenkova:2005ct,Pospelov:2008zw}. $U$-bosons have received much recent attention as a possible implementation of a class of models that could account for the PAMELA positron excess (if it is not astrophysical in origin)~\citep{ArkaniHamed:2008qn,Sundrum:2009}. The idea is that a heavy (hundreds of GeV) dark matter particle $X$ could annihilate into  GeV \zpr\ s, which would decay preferentially into light
SM particles because of its small mass. The necessary large enhancement in the $XX$ annihilation cross section (needed in all such models) could be accounted for by the Sommerfeld enhancement (from the distortion of the wave functions at low energy), due to repeated \zpr\ exchange
 \\
  \hspace*{3cm} \begin{minipage}[t]{5cm} 
\includegraphics[scale=0.6]{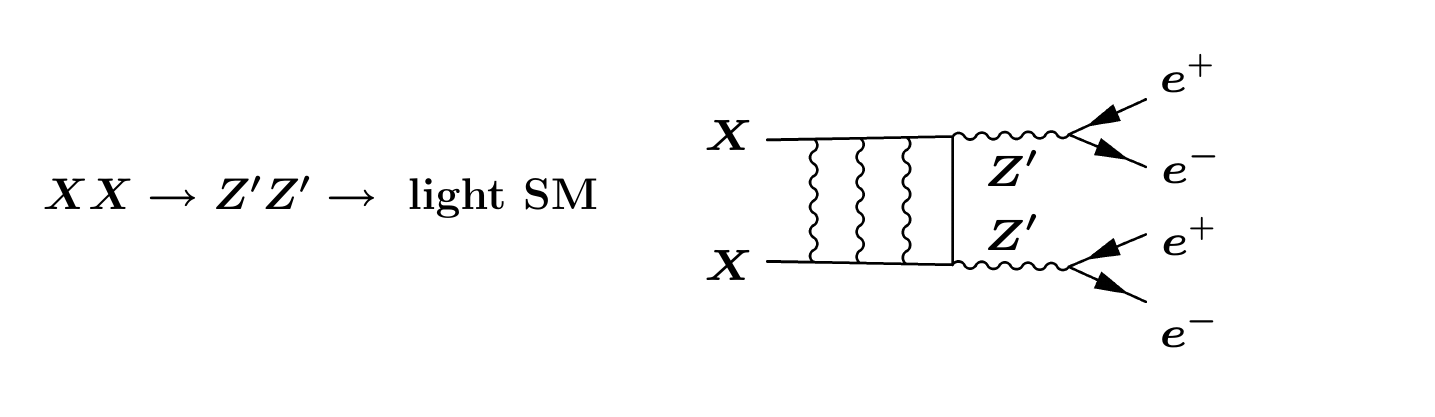}
\end{minipage}\\
The same model could also account for the DAMA results if the \zpr\ couples inelastically~\citep{TuckerSmith:2001hy}, e.g., to
$X_1X_2$ where there is a small (\ord{100 \text{ keV}}) mass splitting between
the $X_1$ and $X_2$.
At the LHC, such a light \zpr\ would be highly boosted, leading to  narrow ``lepton jets''
from $Z' \ra \ell^+ \ell^-$, possible displaced vertices, 
etc.~\citep{ArkaniHamed:2008qp,Gopalakrishna:2009yz,Essig:2009nc,Bjorken:2009mm,Baumgart:2009tn,Cheung:2009su,Reece:2009un,Morrissey:2009ur,Coriano:2008wf}.

\subsection{The St\"uckelberg extension of the Standard Model}
For a \uprm\ (but not a non-abelian theory) it is possible to generate a (St\"uckelberg) \zpr\ mass~\citep{Stueckelberg:1900zz}
without breaking gauge invariance  by writing
\beq \lag=-\frac{1}{4} C^{\mu\nu}C_{\mu\nu}-\oh (mC^\mu+\partial^\mu \sigma) (mC_\mu+\partial_\mu \sigma), \eeql{f13}
where $C_\mu$ is the gauge field which gauge transforms as $\Delta C_\mu = \partial_\mu \beta$, $C^{\mu\nu}$ is the invariant field strength,  and   $\sigma$ is an axion-like scalar, with $\Delta \sigma = -m \beta$. 
The gauge fixing terms cancel the $C_\mu \partial^\mu \sigma$ cross term, leaving a massive vector and a decoupled $\sigma$, i.e., without a physical Higgs boson. Such terms can arise in several ways, 
including in five-dimensional $U(1)$ models, as the  $(\mu^2, \lambda) \ra \infty$ limit of the Higgs model with $(-\mu^2/\lambda)$ fixed; or  in string/brane models  with a Green-Schwarz mechanism~\citep{Nath:2008ch,Antoniadis:2009,Armillis:2008vp}.

Recently, a St\"uckelberg extension of the SM has been proposed~\citep{Kors:2004dx,Feldman:2006ce}, with
\beq
\lag=\lag_{SM}-\frac{1}{4} C^{\mu\nu}C_{\mu\nu}-\oh (m_2C^\mu+ m_1 B^\mu + \partial^\mu \sigma)^2+ g_X C_\mu J^\mu_X,
\eeql{14}
where $B$ is the SM $U(1)_Y$ boson, the new gauge field $C$ couples to a  hidden sector current $ J_X$
with coupling $g_X$, and the ratio $\vp \sim m_1/m_2$ is assumed to be small. The St\"uckelberg term generates a mass $\sim m_2$ for the heavy boson ($\zpr \sim C$). The photon remains massless but couples to the hidden sector with a tiny effective charge $\propto \phi$. Similarly, the  \zpr\ couples to ordinary matter with a suppressed coupling $\propto \phi$. If there are no  open decay channels into the (unsuppressed) hidden sector, this will imply a very narrow \zpr,   e.g., with \mzp\ in the 100 GeV-TeV range. There have been extensions of the model~\citep{Nath:2008ch} to the MSSM and to include kinetic mixing. Dark matter interpretations can involve either  milli-weak or milli-charged particles, and it is possible to account for the PAMELA events.

\subsection{Hidden Valley Models}
Hidden valley models~\citep{Strassler:2006im,Han:2007ae} assume the existence of a strongly-coupled hidden sector which is connected  to
the SM particles by a weakly coupled \zpr\ or other suppressed interactions. Hidden valley neutral  bound state particles
may escape the detector or
may decay back to SM particles with displaced vertices. A  \zpr\ which mainly decays to invisible particles may be detected by associated production,
$pp\ra Z \zpr, \gamma \zpr$~\citep{Gershtein:2008bf}.

\subsection{$Z'$- mediated Supersymmetry Breaking}
\zpr-mediation~\citep{Langacker:2007ac,Verlinde:2007qk} is a top-down motivated scenario, which can occur when a new \uprm\
couples to both the SM sector and the supersymmetry breaking sector, allowing supersymmetry breaking
to communicated by the $M_{Z'}-M_{\tilde Z'}$ mass difference. For example, if the \uprm\ gauge symmetry
is {\em not} broken in the hidden sector, one may still generate a \zpr\ gaugino mass
at a supersymmetry breaking scale $\Lambda_S$ by its (model dependent) couplings to the hidden sector particles. Then scalar mass-squares and $A$ terms for the SM particles can be generated at one loop-level by the
couplings to the \zpr\ and $\tilde Z'$, and the MSSM gaugino masses at two-loop level:\\
\hspace*{1cm}
\begin{minipage}[t]{4.0cm}
\includefigurevh{0.4}{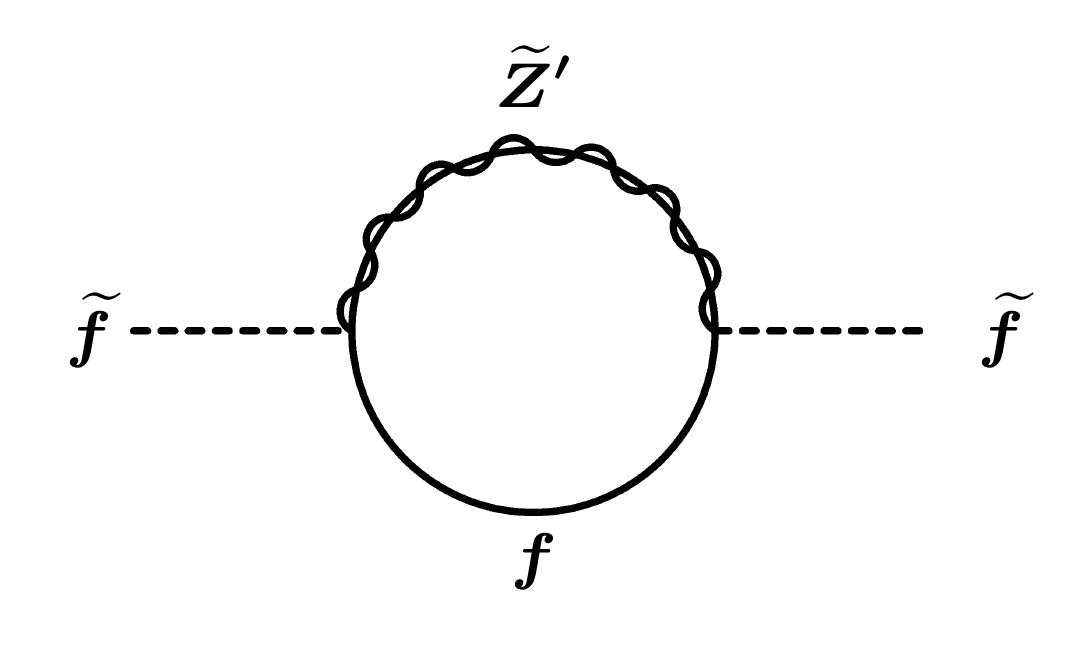}{0cm}{0cm}
\end{minipage}
\begin{minipage}[t]{4.0cm} 
\includefigurevh{0.4}{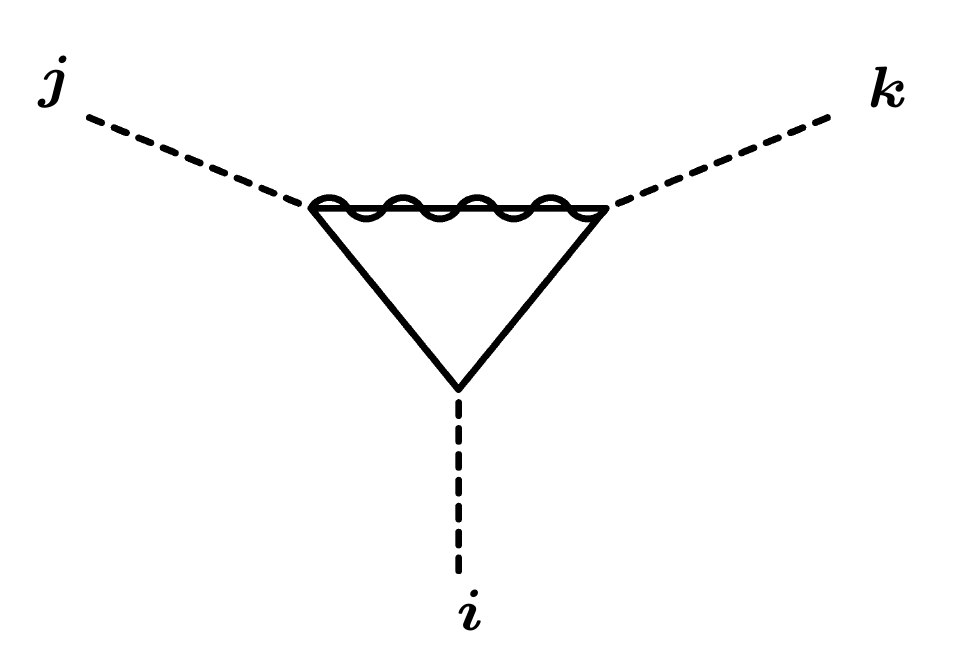}{0cm}{0cm}
\end{minipage}
\begin{minipage}[t]{4.0cm} 
\includefigurevh{0.4}{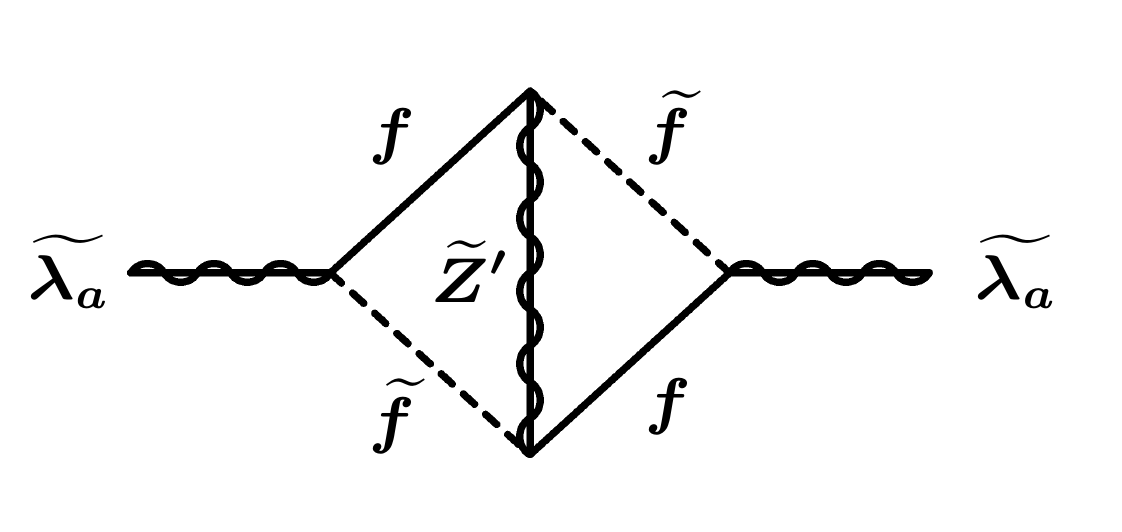}{0cm}{0cm}
\end{minipage}\\
The ratio of the scalar to SM gaugino masses generated in this way is
$m_{\tilde f_i}/M_a\sim (4\pi)^3 \sim 1000$. Since the SM gaugino masses need to be sufficiently heavy,  $M_a\gtrsim 100$ GeV, this suggests two options:
\begin{itemize}
\item The \zpr\ mediation accounts for all of the soft breaking. For $M_a \sim  100-1000$ GeV  one then requires 
scalar masses $\sim 100$ TeV. The electroweak scale must be achieved by a fine-tuning, yielding a
mini version of split supersymmetry (the original version~\citep{ArkaniHamed:2004fb} typically assumed $m_{\tilde f}\sim 10^9$ GeV). Only the gauginos, a single Higgs, and possibly the \zpr\ would be observable at the LHC.
\item The scalar masses could be at 100-1000 GeV. The SM gauginos would have to acquire mass by some other mechanism. For example, the combination with anomaly mediation  may yield a realistic spectrum~\citep{Deblas:2009,Kikuchi:2008gj}.
\end{itemize}

\section{Implications}
Finally, we comment on a TeV-scale \zpr\ with electroweak scale couplings. The implications of such a
discovery would go far beyond the existence of a new vector particle. Here, 
we list some of the possibilities (see~\citep{Langacker:2008yv} for more complete references).

\begin{itemize}
\item There are stringent limits on \mzp\ and on 
$Z-Z'$ mixing from weak neutral current, $Z$-pole, and LEP 2 experiments, and from direct searches by CDF and D0 at the Tevatron, yielding $\mzp > 800-900$ GeV for typical models with electroweak coupling, 
and $|\theta| < \text{ few }\x 10^{-3}$ (e.g.,~\citep{Erler:2009jh}). The LHC has a discovery potential to $\sim 4-5$ TeV through
 $p p \ra \zpr\ra e^+ e^-, \mu^+ \mu^-, jj, \bar b b, \bar t t, e\mu, \tau^+\tau^-$, and should be able to make diagnostic studies of the
 \zpr\ couplings up to $\mzp\sim 2-2.5$ TeV. Possible probes include forward-backward asymmetries; rapidity
 distributions; lineshape variables; associated production of  $\zpr Z, \zpr W, \zpr \gamma$; rare (but
 enhanced) decays such as $\zpr\ra W \bar f_1 f_2$ involving a radiated $W$;
and decays such as   $\zpr\ra W^+ W^-, Zh, 3Z,$ or $W^\pm H^\mp$, in which the small mixing is compensated by the longitudinal $W,Z$ enhancement. (See~\citep{Langacker:2008yv,Rizzo:2006nw,Leike:1998wr} for the classic references, and~\citep{Barger:2006hm,Godfrey:2008vf,Petriello:2008zr,Barger:2009xg} for recent
developments.)

\item A \uprm\ can yield a natural solution to $\mu$ problem in supersymmetry, by forbidding
an elementary $\mu$ term but allowing the superpotential term $W \sim \lambda_S S H_u H_d$, where $S$ is a 
SM singlet charged under the \uprm. This induces an effective $\mu$ parameter $ \mu_{eff} = \lambda_S \langle S \rangle$,
which is usually of the same scale as the soft supersymmetry breaking parameters. This mechanism is similar to the NMSSM (e.g.,~\citep{Maniatis:2009re}), but is automatically free of induced tadpole and domain wall problems.

\item TeV scale \uprm\ models generally involve an extended Higgs sector, requiring at least a SM singlet $S$
to break the \uprm\ symmetry.  New $F$ and $D$-term contributions can relax the theoretical upper
limit of $\sim 130$ GeV on the lightest Higgs scalar in the MSSM up to $\sim 150$ GeV, and
smaller values of $\tan \beta$, e.g. $\sim 1$, become possible. Conversely, doublet-singlet
mixing can allow a Higgs lighter than the direct SM and MSSM limits. Such mixing as well as
the extended neutralino sector can lead to non-standard collider signatures~\citep{Barger:2006dh,Accomando:2006ga}.

\item \uprm\ models also have extended neutralino sectors~\citep{Barger:2007nv}, involving at least the
$\tilde Z'$ gaugino and the $\tilde S$ singlino, allowing  non-standard couplings (e.g., light singlino-dominated),
extended cascades, and modified possibilities for cold dark matter, $g_\mu-2$, etc.

\item Most \uprm\ models (with the exception of those involving $B-L$) require new exotic fermions to cancel
anomalies. These are usually non-chiral with respect to the SM (to avoid precision electroweak constraints) but chiral under the \uprm. A typical example is a pair of \st-singlet colored quarks $D_{L,R}$  with charge $-1/3$.
Such exotics may decay by mixing, although that is often forbidden by $R$-parity. They may also decay by diquark or leptoquark couplings, or they be quasi-stable, decaying by higher-dimensional operators~\citep{Kang:2007ib,King:2005jy}.

\item A heavy  \zpr\ may decay efficiently into sparticles, exotics, etc., constituting a ``SUSY factory''~\citep{Kang:2004bz,Baumgart:2006pa,Cohen:2008ni,Lee:2008cn,Ali:2009md}.

\item The \uprm\ charges may be family non-universal (especially in string constructions), leading to
FCNC when fermion mixings are turned on. The limits from $K$ and $\mu$ decays and interactions are
sufficiently strong  that only the third family is likely to be non-universal~\citep{Langacker:2000ju}. Third family non-universality
may lead to significant tree-level effects~\citep{Barger:2009qs}, e.g., in  $B_s-\bar B_s$ mixing or in charmless $B_d$ decays, competing with SM loops, or with enhanced
loops in the MSSM with large $\tan \beta$.

\item A TeV-scale \uprm\ symmetry places new constraints on neutrino mass generation.
Various versions allow or exclude Type I or II seesaws, extended seesaws, or small Dirac masses by 
higher-dimensional operators~\citep{King:2005jy,Kang:2004ix,King:2009,Mohapatra:2009,Valle:2009}; 
small Dirac masses by non-holomorphic soft terms~\citep{Demir:2007dt}; either Majorana (seesaw) or small Dirac masses by string instantons~\citep{Blumenhagen:2009qh,Cvetic:2009};
and Majorana masses associated with spontaneous $R$-parity violation~\citep{Perez:2009bw}.

\item Electroweak baryogenesis is relatively easy to implement because of the cubic $A$ term 
associated with the effective $\mu$ parameter and possible tree-level $CP$ violation in the Higgs sector (which is not significantly constrained by EDMs)~\citep{Kang:2004pp}.
\end{itemize}

\section{Conclusions}
\begin{itemize}
\item New \zpr\ s are extremely well motivated.
\item \mzp\ at the TeV scale is likely, especially in supersymmetry and alternative EWSB models.
\item The LHC has a discovery potential  to $\sim 4-5$ TeV, and diagnostic capabilities to $\sim 2-2.5$ TeV.
\item There are major implications of a TeV-scale \zpr\  for particle physics and cosmology.
\item A \zpr\ (massless, of at the GeV or TeV scale) could be a portal to a hidden and/or dark  sector.\end{itemize}


\begin{theacknowledgments}
This work was supported by  NSF PHY-0503584 and by the IBM Einstein Fellowship.
\end{theacknowledgments}

\bibliography{pgl}

\end{document}